\documentclass[manuscript,screen,nonacm]{acmart}

\usepackage{booktabs}
\usepackage{multirow}
\usepackage{multicol}

\AtBeginDocument{%
  \providecommand\BibTeX{{%
    \normalfont B\kern-0.5em{\scshape i\kern-0.25em b}\kern-0.8em\TeX}}}

\newcommand{\ignore}[1]{{}}

\newcommand{\new}[1]{#1}
\newcommand{\say}[1]{\emph{``#1''}}

\ignore{
\setcopyright{acmcopyright}
\copyrightyear{2022}
\acmYear{2022}
\acmDOI{10.1145/1122445.1122456}
}

\ignore{
\acmConference[FAccT '22]{ACM Conference on Fairness, Accountability, and Transparency 2022}{June 21--24, 2022}{Seoul, South Korea}
\acmBooktitle{ACM Conference on Fairness, Accountability, and Transparency 2022,
  June 21--24, 2022, Seoul, South Korea}
\acmPrice{15.00}
\acmISBN{978-1-4503-XXXX-X/18/06}
}

\begin{document}

\title{A Field Study of a Human-Centered Process for Increasing AI Transparency}


\author{David Piorkowski}
\email{djp@ibm.com}
\orcid{0000-0002-6740-4902}
\affiliation{%
  \institution{IBM Research AI}
  \city{Yorktown Heights}
  \state{NY}
  \country{USA}
  \postcode{10598}
}

\author{John Richards}
\email{ajtr@us.ibm.com}
\orcid{0000-0001-8489-2170}
\affiliation{%
  \institution{IBM Research AI}
  \city{Yorktown Heights}
  \state{NY}
  \country{USA}
  \postcode{10598}
}

\author{Michael Hind}
\email{hindm@us.ibm.com}
\orcid{0000-0002-7247-7225}
\affiliation{%
  \institution{IBM Research AI}
  \city{Yorktown Heights}
  \state{NY}
  \country{USA}
  \postcode{10598}
}


\begin{abstract}
To address growing concerns about the potential harms of artificial intelligence (AI), societies have begun to demand more transparency about how AI models and systems are created and used. Several efforts have proposed documentation templates containing specific questions to be answered by model developers. These templates provide a useful starting point, but no fixed set of questions can cover the variety of AI use cases and the information needs of diverse documentation consumers. In this paper we evaluate a human-centered process to discover the content needed to effectively document different models and use cases. We report on the experiences of a team in AI for healthcare, none trained in human-centered techniques, that used this process for their own models. Analysis of the benefits and costs of this process are reviewed and suggestions for further improvement in both the process and supporting tools are summarized.
\end{abstract}


%
%
%



\maketitle

\section{Introduction}
AI-based systems are increasingly being used 
to make or inform decisions that can greatly impact an individual's life.
Although these systems can be beneficial, 
\new{research has} \new{shown significant potential for harm from} these systems~\cite{propublica,Selbst2017,amazon-recruiting-2018,UK-exam-2020,zillow-2021}.
Due to these concerns, there has been strong demand for increased AI transparency from researchers, civil society advocates, and
governments.  
One way to \new{improve} transparency is to have a standardized
document, similar in spirit to a nutritional label, that would 
describe the relevant aspects of the AI system.
Several research~\cite{datasheets-cacm-2020,dataset-nut-label-2gen-2020,factsheets-2019,model-cards,aboutml-2020,co-designing-checklists-2020}, industry~\cite{tm-forum-model-datasheet-2020},
and government~\cite{UK-transparency-proposal-2021,EuropeanCommission2020,canada-protocol-2019}
efforts have proposed documentation templates containing questions to be answered by model or dataset developers.
Although these templates provide a useful starting point, 
no single template, 
\new{with a fixed set of questions,}
can cover the \new{variety of AI use cases and the information needs of} diverse documentation consumers.  
These \new{consumers} include people not involved with the development of the AI model: e.g., end users, affected subjects, and regulators, as well as those
who have developed and deployed the model: e.g., data scientists, model validators, and model operations engineers. 

\new{Although}
a universal \new{documentation template} may be \new{useful}, it will leave gaps due to the specifics of the use case, model implementation, problem domain, and regulatory context. To address these gaps, we can strive \new{instead} for \new{an efficient}, repeatable \emph{process} for creating \new{customized}
\new{documentation templates}.
Richards et al. ~\cite{factsheet-methodology-IEEE} 
\new{were the first to propose}
such an approach, following the principles of human-centered \new{methodologies including participatory ~\cite{spinuzzi2005methodology} and value sensitive ~\cite{friedman2013value} design}.
\new{The approach identifies}
the needs of both producers and consumers of AI documentation 
\new{and was initially applied to the creation of customized templates for}
AI FactSheets~\cite{factsheets-2019}.  
Although promising, \new{the approach} has yet to be evaluated with \new{model developers not trained in human-centered techniques}.
\ignore{Richards et al. [25 ] proposed such an approach, following the principles of human-centered design, to
identify the needs of both producers and consumers of AI documentation in the form of a methodology to create AI FactSheets [1].}


\ignore{
\new{CANDIDATE FOR CUTTING ... probably don't need level of detail in the following paragraph quite yet}
This \new{human-centered approach~\cite{factsheet-methodology-IEEE}}
\new{centers} on three roles: content producers, content consumers, and a FactSheet team (FS team). The \emph{FS team} follows the \new{process} and shepherds the creation of a FactSheet \new{template}. During this process, they collaborate with both \emph{producers}, who have the information to be captured, and \emph{consumers}, who have specific documentation needs. The FS team determines consumers' needs and records them as fields to be completed in a FactSheet template. Then, the FS team iterates with producers and consumers to fill out and refine the template \new{along with one or more instantiated FactSheets}.
}

The goal of this paper is to evaluate this \new{process} via a field study of \new{development teams documenting} multiple real AI models and systems over three months. \new{While acknowledging that people outside a development organization (auditors, regulators, affected users) need documentation tailored for them, we focus here on those building models, documenting them, and using this documentation for internal purposes. Without accurate and complete internal documentation, creating external public-facing documentation becomes difficult.}
We interviewed 16 \new{members of these teams} to address the following research questions:
\begin{itemize}
    \item[RQ1] Is the \new{documentation process} usable by team members not trained in human-centered practices?
    \item[RQ2] How well did the resulting documentation address the needs of different consumers?
    \item[RQ3] What did consumers and producers of the documentation see as the benefits and costs \new{of following this process}?
\end{itemize}

\section{Background} \label{sec-background}

AI transparency requires the availability of model and system documentation that is understandable and trustworthy. Documenting any software artifact \textit{well} requires effort. AI models, and the systems in which they are embedded, pose additional documentation challenges \cite{experiences-2020}. First, AI development is highly collaborative. People with diverse skills, specialized vocabularies, and unique tools all contribute to the final deployed model or system. Second, AI development is highly iterative. Model development often begins with a series of lightweight experiments. Multiple threads are followed with frequent false starts and backtracking. Along the way, important decisions about training data manipulation and algorithm refinement are frequently lost. Finally, tooling to support traditional software development is not optimized for AI development and documentation. Proposals for new mechanisms for collecting important facts about AI development throughout the lifecycle~\cite{experiences-2020}, may lead to improved tooling in the future. Currently, only the most disciplined teams can reliably capture information needed for complete and accurate documentation.

\subsection{Documentation Process}
\new{While the human-centered process proposed by Richards et al.~\cite{factsheet-methodology-IEEE} targeted the production of a particular form of AI documentation, FactSheets \cite{factsheets-2019, fs360}, the approach is not limited to this form.
Even so, it is important to discuss FactSheets to appreciate the nature of this documentation and the complexities of producing it.
}
\ignore{To tackle the complexity of creating AI documentation that affords effective transparency, Richards et al.~\cite{factsheet-methodology-IEEE} have proposed a human-centered \new{process} for creating the form of documentation called AI FactSheets. FactSheets \cite{factsheets-2019, experiences-2020, fs360} }
FactSheets
are collections of important facts about the development, testing, and deployment of AI models and systems. These include facts about model purpose, training data selection and cleaning, algorithm selection and tuning, testing for accuracy, bias, privacy risks, adversarial attacks, etc. Different \textit{consumers} of FactSheets (data scientists, business owners, system integrators, deployment engineers, risk officers, regulators, end users, affected subjects, etc.) will have different skills and different information needs. By engaging with these consumers, it is possible to discover both common and unique needs and, thereby, create FactSheets that meet those needs. Those needs may include how to render the documentation, providing for example, a condensed table view for a quick overview or slides to aid presentation of the model~\cite{fs360}. By engaging with fact \textit{producers}, it is also possible to discover what facts they can reasonably create for consumption by others and how to help them create these facts without imposing unacceptable overhead.
\new{Early work with developers of AI models \cite{experiences-2020} exposed the need for substantial flexibility in what facts were important to collect and how those facts might be rendered for different consumers}.

\ignore{
\new{The resulting process for creating FactSheet templates was heavily inspired by prior work in human-centered design methodologies like Participatory Design~\cite{spinuzzi2005methodology} and Value Sensitive Design~\cite{friedman2013value}. Drawing on Participatory Design, the process sought to infuse a spirit of open exploration and discovery with stakeholders, followed by prototyping and iteration.
Drawing on Value Sensitive Design, the process laid out a series of steps to guide the conceptual, empirical, and technical work needed to generate documentation suitable for these stakeholders. The process strives to leverage established human-centered design principles to gently bring these practices into the AI documentation process for an audience, data scientists and software professionals, who are generally unfamiliar with them. TODO condense and move to beginning.} 
}

The \new{process} proposed by Richards et al.~\cite{factsheet-methodology-IEEE} consists of seven steps. The following list presents a somewhat idealized view insofar as there may be iterations within and between steps. In other cases, some of the steps may be collapsed due to informants playing both fact consumer and fact producer roles within the lifecycle~\cite{experiences-2020}. Further details on these steps, including example  questions to explore with informants, can be found in \cite{factsheet-methodology-IEEE}.  
\begin{enumerate}
    \item \textbf{Know Your Documentation Consumers}. AI documentation is produced so that it can be consumed. Understanding the information needs of documentation consumers is the first and most important task. This initial exploration need not be formal. Working with even one representative informant from each major process in the AI lifecycle will provide useful insights into the overall set of consumer needs.  
    \item \textbf{Know Your Documentation Producers}. Some facts to be documented can be automatically generated by tooling. Some facts can only be produced by a knowledgeable human. Both kinds of facts will be considered during this step. Again, working with even one representative producer of relevant facts from each major process in the AI lifecycle will provide the information needed to proceed to the next step.
    \item \textbf{Create a Documentation Template}. What is learned in the first two steps leads directly to the most important part of creating useful documentation, namely the creation of a document template. This template will contain what can be thought of as questions or, alternatively, the fields in a form. Each individual document will contain the answers to these questions. For example a template may start with the question ``What is this model for?''. It may then expand on that question by asking for which uses the model is well-suited and for which the model is ill-suited.
    \item \textbf{Fill In Documentation Template}. This step is where the creator or creators of the documentation template attempt to fill it in for the first time. As this is done, it is important to informally assess the quality of the template itself by reflecting on what was learned about both consumers and producers, their skills, and information needs in the first two steps. While this assessment is not a substitute for further work with them (to follow), it may quickly highlight where improvements are needed.
    \item \textbf{Have Actual Producers Create Documentation}. In this step, actual fact producers fill in the template for their part of the lifecycle.
    \new{In many cases, a producer will be involved in more than one stage (such as model development and model testing)}.
    \ignore{For example, if there is a question in the template about model purpose, find someone who would actually be providing that information and have them answer the question. Ask a data scientist to answer the questions related to the development and testing of the model. If this model was validated, the model validator would enter information about that process. Similarly, a person responsible for model deployment would answer questions related to deployment and ongoing monitoring. If the lifecycle is not that structured, the person responsible for most of the work could fill in this template.}
    \item \textbf{Evaluate Actual Documentation with Consumers}. In this step an assessment is conducted of the quality and completeness of the actual document produced in the previous step. If the document is intended to be used by multiple roles (not uncommon), it should be evaluated separately for each role. To ground this assessment properly, each consumer should be asked to reflect and comment on how this document would actually help them perform their work or provide information to address their concerns.
    \item \textbf{Devise Other Templates for Other Audiences and Purposes}. This step returns to the beginning and is the only \textit{defined} iteration in the \new{process}. There may be other consumers that need to be supported. Or, if the work so far has focused on the process of creating and deploying AI models, it would be worthwhile to consider consumers beyond this such as internal or external review boards or regulators, sales personnel, and end users or others affected by the product or service.
\end{enumerate}

\section{Related Work} \label{sec:related}
Numerous proposals have been made by research groups
~\cite{datasheets-cacm-2020,dataset-nut-label-2gen-2020,factsheets-2019,model-cards,experiences-2020,aboutml-2020,co-designing-checklists-2020}
and government agencies~\cite{UK-transparency-proposal-2021,EuropeanCommission2020,canada-protocol-2019}
for what \new{should be included in AI documentation}.
Several proposals have \new{been shaped by comments from interested parties including industry and individual subjects who might be affected by AI.}
In recent work,
Domagala and Spiro~\cite{UK-transparency-study-2021}
engaged with citizens to better understand their needs regarding information about algorithmic systems deployed by the UK government. This \new{is} one way to approach Step 1 of the \new{process} described in Section~\ref{sec-background}.  
Earlier, Hind et al.~\cite{experiences-2020} explored the needs of documentation producers 
via interviews and a documentation creation exercise lasting a few hours. This paper differs from these works \new{in examining the process of creating AI documentation} in situ through a months-long field study involving both producers and consumers of documentation.

Software documentation, in general, describes important characteristics of a system, application, or module. Despite its necessity, developers and software engineers find the task of \textit{documenting} uninteresting and its creation often falls to technical writers who do not know all the details of the software they are documenting \cite{parnas2011precise}. 
Not surprisingly, consumers of documentation generally dislike what is produced because it is often incomplete, difficult to understand, or out of date \cite{chomal2015software}. 
But documentation is essential as it remains one of the few channels of communication between the developers of  software and its various users \cite{kipyegen2013importance}. 

For conventional software documentation, researchers have developed sets of rules \cite{parnas2011precise}, evaluation critieria \cite{piorkowski2021ai}, and assessment methodologies \cite{smart2002assessing,plosch2014value,ding2014knowledge}. Quality attributes such as accuracy, concreteness, writing style, and understandability have been offered as useful dimensions of quality \cite{plosch2014value}. Other dimensions such as completeness, unambiguity, conciseness, and ease of access have been proposed \cite{parnas2011precise}. Yet other dimensions include consistency, traceability, reusability, format, trustworthiness, and retrievability \cite{ding2014knowledge,zhi2015cost}. This  paper draws on many of these dimensions to assess the quality of AI \new{documentation produced by practitioners employing the human-centered approach outlined above}.

\section{Evaluation Methodology} \label{sec-eval-methodology}
To answer our research questions, we partnered with an AI organization in the healthcare domain that was piloting the \new{documentation process}~\cite{factsheet-methodology-IEEE} \new{to create FactSheets} for several models.
Data for the evaluation reported here consisted of two sets of interviews for each model: one for \new{team members} who followed the \new{process} to create the FactSheet template and associated FactSheets, and one for FactSheet consumers to assess how the resulting FactSheets \new{met} their needs.
Together, these two perspectives provide a \new{somewhat} holistic view of the usefulness of the \new{documentation process}. \new{It should be noted that this methodology relied on participants to recall their experiences, albeit aided with specific prompts outlined below. An alternative approach, where the researchers were embedded with the quite dispersed teams, would have yielded further details but at quite substantial costs and with the risk of altering patterns of work.}

\subsection{Study Participants and Field Study Context}
\new{While there was some overlap between roles, participants in our study generally fell into one of} two roles:  \textit{FactSheet consumers} and \textit{FactSheet (FS) team members}\footnote{The FS team members were generally also the knowledgeable producers of facts in this field study. We ignore the distinction between FS team members and producers in what follows. \new{(This is a good example of how the distinctions in the the documentation process can be adapted to the realities of an actual development organization.)}} \new{In} the FactSheets \new{documentation process}, 
\textit{consumers} are the end users of the created \new{FactSheets}. FactSheet team members are the ones driving the \new{process} for each of the models, defining the FactSheet template, and iteratively \new{refining} the FactSheet. 
In total, we ran 17 interviews with 16 participants: seven FS team interviews and ten FS consumer interviews. One participant was interviewed twice. Participant details and which models each participant discussed are provided in Table \ref{tbl:participants}. Throughout this paper, we identify participants using a three-part code. The first character indicates whether a participant was a FS team member (T) or consumer (C). The second part is a participant identification number. The last character notes the model that the participant produced or consumed. 
For example, `C2-B' indicates the second consumer participant for model B. We interviewed T4 about two models: model B and D. 

\begin{table}
\begin{small}
\caption{The FS team (T\#) and consumer (C\#) participants and their roles for each model. The first character in each column header indicates the model under consideration, referred to as model A, model B, model C, and model D.}
\label{tbl:participants}

\hspace{-4em}
\begin{tabular}{c p{1.2in} c}
\cline{1-2} 
\textbf{A} & \textbf{Role}        \\ 
\cmidrule{1-2}  \cmidrule{1-2} 
T1         & \multicolumn{2}{l}{Data Scientist}   \\ 
T2         & \multicolumn{2}{l}{Lead Software Architect}  \\
C1         & \multicolumn{2}{l}{Domain Expert}          \\
C2         & \multicolumn{2}{l}{Product Manager}        \\ 
C3         & Regulatory          \\ 
C4         & Customer        \\ 
C5         & \multicolumn{2}{l}{Business Analyst}  \\
C6         & Research                  \\
C7         & \multicolumn{2}{l}{Data Scientist}    \\ \cline{1-2}
\end{tabular}
\hspace{2em}
\begin{tabular}{c p{1.2in} c}
\cline{1-2} 
\textbf{B} & \textbf{Role}             \\ 
\cmidrule{1-2}  \cmidrule{1-2} 
T3         & \multicolumn{2}{l}{Data Scientist}       \\ 
T4         & \multicolumn{2}{l}{Data Scientist}      \\
T5         & \multicolumn{2}{l}{Data Scientist Manager} \\
C8         & Consultant         \\ 
C9         & \multicolumn{2}{l}{Database Administrator} \\ 
C10        & Customer       \\ \cline{1-2}
& \\
& \\
& \\
\end{tabular}
\hspace{2em}
\begin{tabular}{c p{1.2in} c}
\cline{1-2} 
\textbf{C} & \textbf{Role}  \\ 
\cmidrule{1-2} \cmidrule{1-2} 
T4         & \multicolumn{2}{l}{Data Scientist}        \\  
\cline{1-2}
& \\
& \\
\cline{1-2}
\textbf{D} & \textbf{Role}  \\ 
\cmidrule{1-2}  \cmidrule{1-2} 
T6 & ML Researcher \\
\cline{1-2}
& \\
& \\
& \\
\end{tabular}
\end{small}
\end{table}

The participants in this field study were part of an AI organization responsible for creating and maintaining several different kinds of models for a variety of health and medical use cases. \new{Participants belonged to the same company as the researchers, but belonged to a different organization.} During the field study period, the \new{participants} worked on documentation for four models (Table \ref{tbl:models}) and completed the FactSheets for models A and B. FactSheets for models C and D were in progress and had at least a first FactSheet draft. Because of this, we were only able to interview consumers for models A and B.
The FS team used the publicly available resources on the AI FactSheets 360 site~\cite{fs360} to help them follow the \new{process}. Resources included explanations of what FactSheets are and their intended applicability, a description of the \new{process} for creating FactSheets, example FactSheets, and links to research papers, videos, and a Slack community.

\begin{table}
\begin{small}
\caption{The four models documented during the field study.}
\label{tbl:models}
\begin{tabular}{c p{5in}}
\toprule
\textbf{Model ID} &  \textbf{Model Description} \\ \midrule
A            & 
A set of models to assist Medicaid Fraud, Waste, and Abuse investigators in retrieving relevant information from Medical Insurance policy
                                                                     \\
B            & 
A mature model (version 21) to predict relative mortality risk of a hospitalization stay  \\
C            & 
A model (in development) to identify potential risks of surgical complications to proactively mitigate these potential issues
\\
D            & 
A model to identify populations that are likely to have improved health outcomes if their social factors are improved
                                                       \\ \bottomrule
\end{tabular}
\end{small}
\end{table}

\subsection{Interview Protocols}

For \new{each of the members of the} FS team, we conducted approximately one-hour, semi-structured interviews that focused on how 
the \new{process} worked in practice \new{focusing} on five main topics: (1) the context of, and motivations for, creating the FactSheet, 
(2) how they used the available resources, (3) how they \new{carried out} the FactSheets \new{documentation process,} 
4) perceived benefits of the process, and (5) perceived costs. 

Interviews with the consumers lasted approximately 30 minutes. Unlike the FS team interviews (that focused mainly on the process of creating a FactSheet), consumer interviews focused on whether the created FactSheet met their needs, and if not, where additional or different content was needed. Topics in consumer interviews focused on: (1) the quality of the created FactSheet and (2) how the FactSheet \new{differed from AI documentation they had encountered in the past.}
\new{Interviews were held within two weeks of the completion of each FactSheet for models A and B, or within one week of the completion of the first draft for models C and D. Throughout the interviews, we asked participants to contextualize their responses with respect to their current documentation practices and encouraged participants to share both positive and negative sentiments.}

Both FS team and consumer interviews were similarly run. In each interview, participants were given access to the finished (or in progress) FactSheet that they could refer to as needed. Interview sessions were run remotely, recorded, and transcribed for later analysis. 

\subsection{Data Analysis}
To analyze participants' responses, we used two approaches. For questions that expected more direct responses such as identifying missing content in the FactSheet, we identified and aggregated participants' answers. For more complex questions such as the ones around perceived benefits and costs, we conducted thematic analysis~\cite{braun2006using}, focusing on themes that addressed our research questions. First, participants' responses were extracted from each interview transcript such that a complete response was available for each question, including any relevant context. One author extracted responses, annotating any references a participant made to the FactSheet as necessary.

All authors participated in the thematic analysis. The thematic analysis was an iterative and inductive process, with new themes emerging and collapsing as the authors worked through the data. Using the set of interview responses, each of the authors coded any sentences with a theme or topic relevant to the research or interview questions. Extracted codes and their definitions were iteratively added, consolidated, and removed as needed to form the codes. Upon completion, all three authors together discussed codes and consolidated codes into themes and subthemes.

\section{Results} \label{sec-results}
Our results are structured by research question. We address whether and how the \new{process} was usable by a real-world \new{development team not trained in human-centered practices} (RQ1), if and how the resulting \new{documentation} met the needs of their various consumers (RQ2),
and what the benefits and costs of this approach were seen to be (RQ3).

\subsection{RQ1: Usability of \new{Documentation Process for} FS Team Members}
Practitioners skilled in \new{human}-centered practices will recognize that the \new{process} applies these practices to the creation of FactSheet templates (and the resulting FactSheets). Is it reasonable to expect data scientists, engineers, and others \textit{not} trained in these practices to execute the \new{process} and derive the expected benefits? In this section we report on how well FS team members were able to apply the \new{process}. Each subsection looks at a different aspect of the \new{process}'s usability: the value of available educational resources; how well individual steps of the \new{process} could be followed and how useful they were; and the ways in which the \new{process} helped in documenting AI models.

\subsubsection{How FactSheet Teams Used Educational Resources} \label{sec:education-resources}
The FactSheets 360 website \cite{fs360} provides an overview of the methdology, a number of illustrative FactSheets, and links to more detailed instructional content. When we asked the four FS team members who used the website to identify the resources that were the most useful to them, responses included the example FactSheets on the website (T1-A and T5-B) and the \new{process} summary (T2-A). 
When asked for more details, T1-A described how one example showed how disparate kinds of documentation could be brought together into a single place saying \say{It was self-contained. It provided all the info I needed to understand.} T5-B described how the different views in the example FactSheets, such as the table view and slide view, helped him understand how FactSheets could be tailored to the needs of different consumers saying
\say{Something clicked (for me) on one of the (example) pages about how the same facts can be represented in different ways, which was the key idea for me to really see the value of this approach. I haven't thought about documentation in that way before}.
T6-D echoed this sentiment explaining how many of the discussions around what to document revolved around this perspective on documentation saying \say{There were two critical things on my mind. It was understanding the players. What do we mean by producers? What do we mean by consumers?} By framing documentation as something to be used by others, FS teams changed their orientation from just reporting on what a model did and how it was developed to focusing on how to make the description useful for specific consumers. More details on this observation follow in Section \ref{sec:methodology-pros-cons}.

FS team members did not report that any resources were \textit{un}helpful.  They did identify some gaps and stated that there was too much content to digest it all. T1-A identified procedural guidance that he would have liked to have such as a \say{minimum set of fields} required for a FactSheet template.
T2-A described a desire to have concrete guidance for how long the \new{documentation process} should take and which facts to start with. 
To address this, T2-A developed a schedule based on his team's first attempt at creating a FactSheet template, which provided guidance for subsequent efforts. \say{We needed something to explain the process... Starting with the \new{[documentation process on]} the site and then breaking it out into 'Here's a six-week schedule' basically. To give (the rest of the team) something concrete.} 

\subsubsection{\new{FactSheets Documentation Process} Steps}
\label{sec:methodology-pros-cons}
To better understand how well the individual steps of the \new{process} could be applied, we asked FS team members to reflect on the steps one by one. FS team members responded most often that Step 1 of the \new{process}, \textit{Know your FactSheet Consumers},
was the most useful of the seven steps.
Understanding their consumers encouraged FS teams to consider who they would be writing for, and what
their documentation for them should include. 
Half the FS team members (T1-A, T2-A, T5-B) decided this step was the most important one.
T1-A said, \say{Knowing your consumers was a step that was particularly useful for me... Once I had a particular consumer identified I realize that there are things that I need to think more about, or I need to reach out to people who know about this.} 
Revealingly, T1-A first tried skipping Step 1 to save time, opting to go directly to getting feedback from consumers on a FactSheet based on an example in the FactSheets 360 website. He recalled, \say{
We started filling it out without the right [consumer needs]... and then we kind of realized a lot of this stuff isn't right. So we went back and rewrote quite a lot of it basically.} 
After this setback, the other FS team members decided to speak with potential consumers first.

Participants T2-A and T5-B noted the benefit of creating a first version of a FactSheet (Step 4) before gathering feedback from consumers on just the template (produced in Step 3). While creating this first version might be seen as adding unnecessary time, the conversations with consumers focused on making changes to this version instead of generating content from scratch. 
Since one of largest costs reported by participants in the \new{documentation process} was validating the FactSheet with consumers, FS teams appreciated this time-saver. 
T6-D noted the value of this step as well and added detail about how the act of filling out the FactSheet template spurred reflection on aspects of the model they previously had not considered saying, \say{It's forcing us to answer questions about our model that we may have thought about, but never documented in any way.} 

Although FS team members did not consider any specific steps of the \new{process} to be \textit{un}helpful, they did adapt the \new{process} to better fit their team's needs and timeline. 
FS teams chose, for example, to minimize the seeming seriality of Steps 5 (have actual producers create a FactSheet) and 6 (evaluate the FactSheet with actual consumers). They preferred a more iterative approach where FactSheet content was evaluated as soon as there was enough for a specific consumer role, even if the rest of the information was not yet ready. These changes were implemented during the first pilot. Since the subsequent pilots were informed from the first, T3-B, T4-B, T5-B, T4-C and T6-D all followed the altered protocol, further compressing the timeline by starting with the template from model A.

\begin{table}
\begin{small}
\caption{FS team member responses to \new{whether the process} helped them in various ways. A response of `n/a' indicates that the FS team member was unsure or unable to give a response.}
\label{tbl:methodology-useful}
\begin{tabular}{llllllll|c}
\toprule
\textbf{Did the \new{process} help...}                    & \textbf{T1-A} & \textbf{T2-A} & \textbf{T3-B} & \textbf{T4-B} & \textbf{T5-B} & \textbf{T4-C} & \textbf{T6-D} & \textbf{\# Yes}\\ \midrule
... Improve documentation consistency?        & Yes  & Yes  & Yes  & Yes  & Yes  & Yes  & Yes & 7 \\ 
... Evaluate documentation usefulness?         & Yes  & n/a  & Yes  & Yes  & Yes  & Yes  & Yes & 6 \\
... Identify new documentation needs?          & Yes  & Yes  & No   & No   & Yes  & Yes  & Yes & 5 \\
... Facilitate creating useful documentation? & Yes  & No   & n/a  & No   & Yes  & Yes  & Yes & 4 \\
... Improve documentation practices?           & Yes  & n/a  & Yes  & No   & No   & n/a  & Yes & 3 \\
... Identify new consumers?                    & Yes  & Yes  & No   & No   & No   & n/a  & No  & 2 \\ \bottomrule

\end{tabular}
\end{small}
\end{table}

\subsubsection{How the \new{Process} Helped FactSheet Teams Document AI Models}
FS team members noted how the \new{documentation process} provided specific benefits over their previous documentation practices. Table \ref{tbl:methodology-useful} summarizes the questions about specific benefits that we asked about and participants' responses. All FS team members agreed that the \new{process} would likely enhance documentation consistency, primarily by consolidating what was previously scattered in multiple documents into a single place.
T5-B summarized, \say{We'll be able to consolidate in ways that make sense so that there's one place for facts, rather than, lots of places to keep things up to date.}

T5-B and T6-D described how 
completing the FactSheet template 
had caused them to reflect on the usefulness of what they were writing in the FactSheet fields. 
T5-B described how a consumer need for making the problem description 
reusable
encouraged conciseness. He said, \say{One of the best (consumer) feedbacks that we received was in the way we tried to condense the problem description in the FactSheet so that they could copy and paste it to engage with customers}.

Importantly, five of the FS team members agreed that the \new{process} helped their team identify \textit{new} documentation needs. They discovered several additional template fields as a direct result. 
Table~\ref{tbl:template} shows the non-proprietary fields (most of the fields were non-proprietary) from the FactSheet template for model A.  Newly-discovered fields are italicized.
Examples included regulatory requirements, model maturity, usage considerations, 
and run time requirements. 
T6-D summarized the value stating, \say{There was a lot of information here that we wouldn't perhaps otherwise made available or collected or even thought of.}

\begin{table}
\begin{small}

\caption{The template created by the FS team for model A. Italicized fields indicate new fields that were added during the field study.} 
\label{tbl:template}

    \begin{tabular}{p{2.2in} p{3.45in}}
    \toprule
    \multicolumn{2}{c}{\textbf{Overview}}  \\ \midrule
    Model Name          & The  name of the model                                         \\
    \textit{Version and Dates}   & The model's version and date it was created                    \\
    Problem Solved      & A description of the problem the model is designed to solve    \\
    \textit{Maturity Ratings}    & A evaluation of the model's maturity along multiple dimensions \\
    \textit{Contact Information} & Who to contact about this model                                \\ \toprule
    \multicolumn{2}{c}{\textbf{Intended Use}}  \\ \midrule
    Intended Domain    & The industry of functional area the model intends to support                             \\
    \textit{Use Case}         & The use scenario, including who, what, when and how the model outputs intends to support in practice \\ \toprule
    \multicolumn{2}{c}{\textbf{Data and Model}}  \\ \midrule
    Training Data                        & Statistics and details about training data        \\
    \textit{Cohort / Therapeutic Area Definition} & Population cohorts of interest for this use case  \\
    Model Information                    & Description of the model                          \\
    Input/Output                         & Description of expected model inputs and outputs  \\
    Internal Test Data                   & Description of the internal model test data used  \\
    External Validation Data             & Description of the external model validation data \\ \toprule
    \multicolumn{2}{c}{\textbf{Performance}}  \\ \midrule
    Performance Expectation \& Robustness    & Expected performance metrics and evaluation of the model's adversarial robustness     \\
    Model Performance                        & Description of model evaluation and actual performance metrics                        \\
    Bias \& Fairness                         & Description and evaluation of the model's fairness and bias                           \\
    \textit{Explainability}                           & Description of the model's ability to explain its predictions                         \\ \toprule
    \multicolumn{2}{c}{\textbf{Usage Considerations}}  \\ \midrule
    Optimal \& Poor Conditions               & Conditions under where the model performs well and poorly                             \\
    \textit{Volume \& Latency}                       & Expected number of predictions and time per prediction                                \\
    \textit{Discontinue Use If}                       & Conditions under which to stop using the model                                        \\
    \textit{Runtime Data Requirements}                & Data prerequisites necessary to run this model                                        \\
    \textit{Runtime Technology Requirements}          & Technology prerequisites necessary to run this model                                  \\
    \textit{Use-Case Tolerance of Error}              & Evaluation of the error tolerance of this model                                   \\
    \textit{Market Differentiator}                   & Description of market differentiators for this model                                  \\
    \textit{Related Areas Where Models May Be Useful} & Other potential use cases for this model              \\ 
    \bottomrule
    \end{tabular}
\end{small}
\end{table}

\subsection{RQ2: \new{Documentation} Usefulness for Consumers}

To determine
whether the FactSheets generated by the FS teams were useful to consumers, we asked consumers to 
(1) evaluate the FactSheet's quality along several dimensions and describe any gaps
and (2) to 
describe if and how the FactSheet differed from the AI documentation they had encountered previously.

\subsubsection{Evaluation of Usefulness}

\begin{figure}
    \centering
    \includegraphics[width=1.0\textwidth]{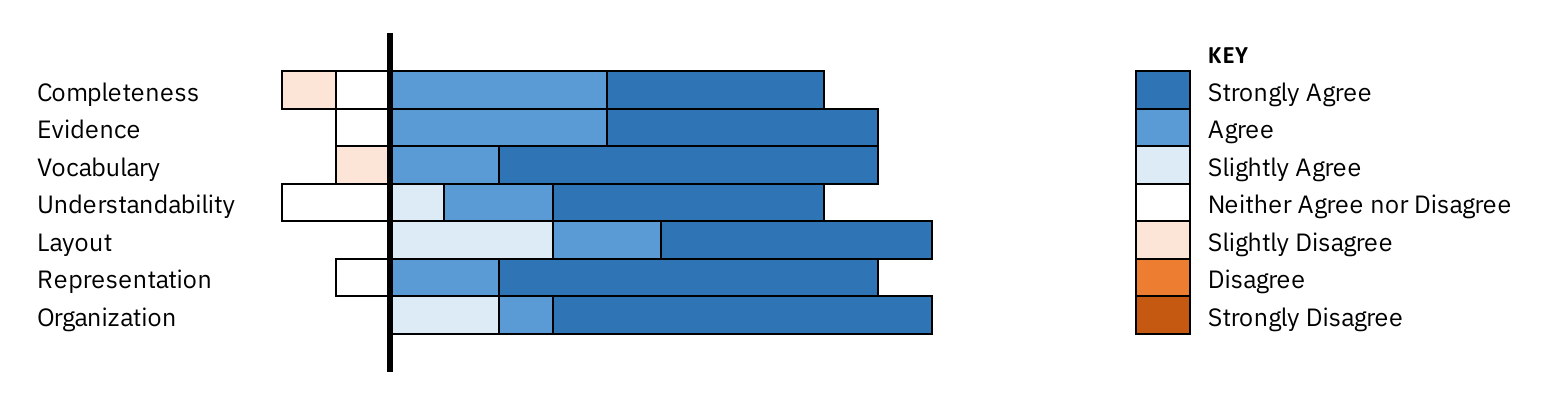}
    \caption{Consumer responses for evaluating the FactSheet quality. The responses on the right side of the dark vertical line represent responses of `slightly agree' or higher.}
    \label{fig:factsheet-quality}
\end{figure}

We asked consumers to evaluate the FactSheets, along multiple quality dimensions, on a seven-point Likert scale ranging from "strongly disagree" to "strongly agree". The dimensions were drawn from work on assessing AI documentation quality \cite{piorkowski2020towards} which adapted concepts of completeness~\cite{parnas2011precise,smart2002assessing}, accuracy~\cite{parnas2011precise,plosch2014value,smart2002assessing}, and conciseness~\cite{ding2014knowledge} from software documentation along with prior research on the pragmatics of effective communication, \new{most notably \cite{grice}}.
The dimensions and prompts are shown in Table~\ref{tbl:quality-prompts}.
We found that the overall quality of the FactSheets produced was excellent, scoring high marks across all the quality dimensions as shown in Figure \ref{fig:factsheet-quality}. These results suggest that the work of the FS teams, perhaps especially the iterative evaluation \new{and refinement }of FactSheet content with consumers, paid off and resulted in relevant, useful, and usable documentation.

\begin{table}
\begin{small}
\caption{Likert-scale prompts given to consumers to assess FactSheet quality.}
\label{tbl:quality-prompts}
\begin{tabular}{p{1.2in} p{4.05in}}
\toprule
\textbf{Quality Dimension} & \textbf{Prompt}                                                                                   \\ \midrule
Completeness      & The FactSheet has all the information that I require for my use case.                                          \\ 
Evidence          & The FactSheet's information is well supported with additional evidence provided where needed.                          \\ 
Vocabulary        & The FactSheet is written using appropriate vocabulary and word choice.                                        \\ 
Understandability & The FactSheet's content and information is easy to understand.                                                    \\ 
Layout            & The FactSheet's structure and layout is intuitive.                                                                \\ 
Representation    & The FactSheet's information is presented in the expected way with text, tables, and figures appropriately chosen. \\
Organization      & The FactSheet's information was well organized and easy to locate.                                                \\ \bottomrule
\end{tabular}
\end{small}
\end{table}

\subsubsection{Things That Were Missing} \label{sec:missing}

\begin{table}
\begin{small}
\caption{Summary of missing documentation information per consumer. Italicized entries were mentioned by multiple consumers. Although market differentiation content existed in the FactSheets, participants wanted \textit{more} detail than what was available, so it is included \new{here}.}
\label{tbl:missing-facts}
\begin{tabular}{c p{1.2in} p{3.7in}}
\toprule
\textbf{Consumer} & \textbf{Role}          & \textbf{Missing documentation fields}                                                                                                                                                                         \\ \midrule
C1-A                 & Domain Expert          & None                                                                                                                                                                                                          \\
C2-A                 & Product Manager        & \textit{Market differentiation}; \textit{existing deployments}; customers; pricing   information; model maturity; links to marketing materials, presentations, and   demos                                                      \\
C3-A                 & Regulatory             & \textit{Market differentiation}; Process model is replacing; benefits and costs of deployment; access   information; model evaluation summary; known risks; applied risk mitigations;   \textit{how to interpret model metrics}; productization details \\
C4-A                 & Customer-Facing               & \textit{Market differentiation}; Mapping between model metrics to business use case; \textit{how to interpret model metrics}; \textit{good/bad ranges for model metrics}; business context                                                                     \\
C5-A                 & Business Analyst       & \textit{Good/bad ranges for model metrics}                                                                                                                                                                             \\
C6-A                 & Research               & Data sources; data cleaning details; data limitations and constraints;   model training details; background of domain experts; definitions for bias;   definitions for explainability                         \\
C7-A                 & Data Scientist         & \textit{Existing deployments}; opportunities for model improvement; customer   feedback; history of model evaluations                                                                                                  \\
C8-B                 & Consultant             & None                                                                                                                                                                                                          \\
C9-B                 & Database Administrator & None                                                                                                                                                                                                          \\
C10-B                & Customer-Facing               & Customer-specific model evaluation                                                                                                                                                                            \\ \bottomrule
\end{tabular}
\end{small}
\end{table}

Even with the high quality-dimension scores, consumers were able to identify things that were still missing. For any response to a prompt with a score of less than `strongly agree', we asked the respondent what was missing or needed to be changed for that dimension. Table \ref{tbl:missing-facts} summarizes their replies.

A closer look at this reveals that missing content tends to be
quite role-specific. 
For example, C2-A's suggestions are related to how the model fits into the larger business context. He wanted to see more information about which customers \new{were} using the model along with pointers to additional materials useful in discussions with potential customers. He stated, \say{It might be good to include 
something about where has the model been deployed?}
The requests from C3-A, whose role involved regulatory concerns, focused on risk: the model context, risk evaluations, and risk mitigation. 
C4-A and C10-B, both in customer-facing roles, focused on how to determine model quality and how the quality compares to competitors. 
C5-A, like C4-A, not having a deep data science background, wanted more guidance on how to interpret the model evaluation metrics, both in terms of how the model relates to the business, but also including the reasonable question of what makes a good metric score. P5-A said, \say{I think I understand what these (metric) scores might tell me, but I have to do a lot of thinking to connect it back to the business problem... What do they mean for me and the situation that I'm trying to evaluate? ... What's a good range? What's a bad range?} 
C7-A, a data scientist, wanted to see information that would help him improve the model, along with a history of model evaluations over time. 
These consumers' role-specific requests underscore the point that different consumers have markedly different documentation needs, and provide further evidence that a one-size-fits-all solution for AI documentation will not suffice.

The combination of high quality scores for the \new{FactSheets} along with several suggestions for further improvement tell a somewhat mixed story. On the one hand, as indicated by the high scores, participants seemed to be satisfied with the content of the FactSheets. On the other hand, they still had suggestions for how to improve them. There are potentially several reasons for this. FS teams may not have iterated enough with consumers to gather all the relevant documentation needs (on average going through only two iterations). Or FS teams intentionally did not address all these needs to keep FactSheet size manageable (a role-based filtering mechanism, see \ref{sec:tools}, may help manage this trade off). Another interesting possibility is that by asking consumers in our interviews to reflect on a FactSheet from the perspective of the quality dimensions, gaps surfaced that they did not see before. If true, this latter interpretation may have implications for eliciting consumer needs more completely.

\subsection{RQ3: \new{Documentation Process} Benefits and Costs}

To better understand the value proposition of both the \new{documentation process} and FactSheets themselves, we asked FS team members and consumers what specific pros and cons they experienced, further reflecting on how this compared with their documentation experiences from before. 

\subsubsection{\new{Documentation Process} Benefits}
As mentioned earlier, one key benefit of FactSheets was to consolidate disparate documentation.
T1-A expressed how bringing in the different viewpoints gave a more holistic picture of the model and its context. He said, \say{When [you] have the document in front of [you] from all points of view, you start joining things together and asking questions... It gives a broader overall picture and fills gaps in your own knowledge.} This broader context would be difficult to capture without input from all the consumers.
Another benefit of consolidation was how the FactSheet facilitated additional exploration. T1-A, T3-B and T6-D discussed how the more broadly written content of the FactSheet coupled with its pointers to where more details can be found enabled such exploration. T1-A said, \say{I think this is useful documentation for the reason that everyone can go in and find what they are looking for, maybe not at the level of detail that [they need], but there is links and hyperlinks to everything.} T6-D referred to the FactSheet as a \say{one-stop shop for the model}. Seven consumers (C2-A, C3-A, C4-A, C5-A, C6-A, C8-B and C9-B) agreed that the FactSheet acted as an effective point of entry for understanding the aspects of the model relevant to their role.

Consumers elaborated on the kinds of further exploration that the FactSheet enabled. For example, C2-A and C8-B explained how the FactSheet eased navigation to other relevant documents. C2-A said, \say{It's a single document that makes it easy for me to navigate to the next level. (Previously), I'd have to hunt down multiple documents or ask people what was available... This seems to bring it together at least in a single launch point where I can read a bit of detail about it, and then I can understand where to go to find more information.} C3-A and C6-A explained that the FactSheet served as a pointer to the right people to contact for further follow up, potentially reducing the time spent gathering information about a model. C3-A summarized, \say{I think it reduces the amount of time that will be spent of clarifying issues... It would allow me to do pre-work without having to do 3 meetings with a group of 5 to 6 people.} C5-A and C7-A likewise agreed that it reduced the number of meetings required to understand the current state of a model. C1-A, C4-A and C10-B echoed similar sentiments for explaining the model to customers interested in purchasing the model.

In addition to serving as a jumping-off point, FactSheets acting as the sole source (or at least sole anchor) of truth, enabled FS teams and consumers to reduce work in several ways. 
Previously,
there would be several sources of documentation that sometimes provided conflicting information. The FactSheet as the single source of truth allowed content to be reliably copied for other creations such as customer-facing descriptions, reports, or presentations. C8-B recalled a time that she was able to reuse content in the FactSheet for a client. She said, \say{This is actually really nice that they've got this [documentation field] added in here because now I can lift this paragraph and put it into a client document... And in an approved definition or language for how to describe [it].} One FS team member, T3-B, likewise used the FactSheet as a text source for additional materials. Simply knowing where the most current documentation is also enabled more effective employee on boarding, especially when the people who worked on the model were no longer available. C6-A summarized, \say{You would talk to [someone] who'd go talk to someone else who has a link somewhere, or sends you some old weird spreadsheet that says, `go talk to this person'. And then someone would have left the company and then they don't have access to something. So this is much much better, because it puts everything in one place.} Finally, FS team members also reported benefits centered around the theme of improved organization. They reported how FactSheets helped them get a better understanding of documentation needs earlier (T1-A, T2-A), reduce repetition (T2-A, T6-D), improve awareness of other documentation (T3-B), and help identify overlaps in existing documentation (T4-C, T5-B).

\subsubsection{\new{Documentation Process} Costs}

\begin{table}
\begin{small}
\caption{The number of hours each FS team member spent working on the \new{template and FactSheet.}}
\label{tbl:hours-spent}
\begin{tabular}{ccccccc}
\toprule
\textbf{T1-A}     & \textbf{T2-A}     & \textbf{T3-B}     & \textbf{T4-B}    & \textbf{T5-B}     & \textbf{T4-C}     & \textbf{T6-D}      \\ \midrule
15 hours & 24 hours & 10 hours & 8 hours & 15 hours & 24 hours & 6-8 hours \\ \bottomrule
\end{tabular}
\end{small}
\end{table}

The most obvious cost faced by participants was simply the time required to create the template and FactSheet. FS team members reported times ranging from 6 to 24 hours (Table \ref{tbl:hours-spent}). This estimate includes all the steps of the \new{process}, including meetings with others and time spent \new{researching and }filling out the fields. Of course, creating documentation of any sort takes time, and it is not clear that this is outside the norm. Scheduling time with consumers was the most common cost mentioned by FS team members, specifically, T1-A, T2-A, T5-B, and T6-D.

Aside from the challenge of getting people together, some steps of the \new{process} required additional effort from the FS team. One mentioned by FS team members T1-A, T3-B and T6-D was Step 4, filling in the FactSheet template for the first round of consumer feedback.
This step is meant to improve the FactSheet template and, as noted above, also made subsequent interactions with consumers more efficient as they could critique something that exists instead of creating it on the spot. However, this approach passed the burden off to the FS team, leaving them to fill out fields they may have been unfamiliar with.

Another challenge in preparing that first draft was locating some of the documentation that existed from past work on the model. Both T3-B and T6-D were working with a models that had scattered documentation. T6-D described how much of her time was spent just finding the source that had the information she needed for the FactSheet. She said, \say{(The difficulty) is not knowing where to look, right? It's reading through documents that you potentially didn't have to read through.}

Other than the missing fields, noted above, 
consumers reported few specific costs to the \new{process}. The one exception was consumer C10-B who did not see the benefit of a FactSheet over the documentation that already existed for model B (with which she was already familiar). C10-B may be a bit of an outlier, however, as all nine of the other consumers found FactSheets to be better than their prior documentation.

\section{Discussion} \label{sec-discussion}
The current field study looks at the early process of adopting the \new{FactSheet documentation process} within an organization. It examines whether the \new{process} can be followed by those with no particular training in human-centered design, and whether the near-term benefits outweigh the costs. We believe this early picture is promising and expect even more gains from long-term adoption due to reuse of templates, growing awareness of what facts matter to consumers, and increased automation of fact collection. 

\subsection{Enabling Reuse} 
Despite our belief that there is no universal template or checklist for AI transparency, there is a strong and quite understandable desire to reuse an existing template, for example, one from \cite{fs360}, rather than developing one from scratch. What we have observed in the present work is that a core set of facts did apply in a very different domain from the ones considered \new{previously}. These included descriptions of model purpose, training data, and model inputs and outputs. As shown in Table \ref{tbl:template}, this core set was augmented with domain-specific facts, such as therapeutic-area definition (for these models in the healthcare space) and market differentiator for models that reached a level of maturity where they are competing for sales within a segment. This kind of "base plus extensions" form of reuse is likely to become a dominant form.

An additional observation from the present work is that once a template has been tailored for a particular organization, it can be used with very little additional work by others within that organization. While the models in this study were all in the healthcare space, they had very different types, use cases, and maturities. Even so, the template developed for the first model was readily reused for subsequent models.

\subsection{Design Implications for Tools}
\label{sec:tools}

One of the key gaps identified by FS team participants was the lack of tool support for gathering facts. Facts needed to be found in existing documentation (large, dispersed, and created for a variety of purposes for a variety of audiences), then adapted for use within the much more compact FactSheet, or were obtained by tracking down and talking with others in the organization. We believe \new{that tooling can improve this situation in several ways:}

\begin{itemize}
    \item Fact automation: Many kinds of facts can be captured automatically by suitably instrumented tools. Code repositories adapted for use in AI development can record key facts about training data and model versioning. Lightweight tools such as Python notebooks can include fact-capture mechanisms with very little overhead. As the field identifies the kinds of facts that best support transparency, this sort of automation will likely appear.
    \item Fact elicitation: Some facts, such as model purpose or non-quantifiable aspects of potential bias, must be entered by a knowledgeable human. A good library of examples, wizards, and other elicitation tools will make this both more efficient and help make facts more consistent (supporting comparison) and consumable.
    \item Fact explanations: There will always be a wide range of expertise among the various consumers of \new{AI documentation}. Automatic mechanisms for tailoring fact content to skill level are not infeasible. In addition, a library of reusable links to definitions of terms and supporting material will allow consumers to better understand potentially complex aspects of AI without requiring fact producers to author this supplemental information repeatedly.
    \item Fact filtering: Either through customized templates or filtering controls, consumers will eventually be able to customize documentation to their needs.
\end{itemize}

\section{Conclusion} \label{sec-conclusions}
Our results suggest that the \new{AI documentation process}~\cite{factsheet-methodology-IEEE} worked as intended, that is, FS team members were able to successfully use the \new{process} to create documentation that was useful to its intended audience. Specifically, we found \new{the following answers to our research questions:}
\begin{enumerate}
    \item[RQ1] \textbf{\new{Process} usable without human-centered training?} Even without training in human-centered practices, FS team  members adopted the \new{process}, successfully elicited needs from their consumers, and encoded those needs into a FactSheet template. Furthermore, the template addressed needs specific to the AI healthcare domain, but was still general enough to be reused across several, quite different, models.
    \item[RQ2] \textbf{Documentation addressed needs of consumers?} Consumers found the resulting FactSheets met their needs, giving them high quality scores along several dimensions. Not all needs were met, however, and the identification of missing content further reinforced a core motivation for the \new{documentation process}: that AI documentation must be tailored to the specific needs of each consumer.
    \item[RQ3] \textbf{\new{Process} benefits and costs?} FS team members and consumers benefited from documentation that was authoritative, in a single location, and supported additional detailed exploration for specific needs. In 16 of 17 interviews, participants agreed that the FactSheets were an improvement over earlier documentation practices.
\end{enumerate}

In summary, the benefits of following the \new{FactSheets documentation process} to improve AI transparency seemed to outweigh the costs, and are best expressed by T2-A who closed their interview saying the following about the \new{process}, \say{I think it's going to be very useful for us going forward and really it's becoming a core part of what we're doing.}

\ignore{
\begin{acks}

\end{acks}
}

\bibliographystyle{ACM-Reference-Format}
\bibliography{main}

\ignore{
\appendix

\section{Appendix section} \label{sec-appendix}

\subsection{Appendix subsection}

Do we have anything to include in the Appendix?
}

\end{document}